\documentclass[letterpaper, 10 pt, conference]{ieeeconf}  
\IEEEoverridecommandlockouts    

\usepackage{graphics} 
\usepackage{epsfig} 
\usepackage{subfig}
\usepackage{stfloats} 
\usepackage{float}
\usepackage{amsmath} 
\usepackage{amssymb}  
\usepackage{color} 
\usepackage{siunitx}
\usepackage{multirow}
\usepackage{hyperref}
\usepackage{booktabs}
\usepackage{footnote}
\usepackage{tablefootnote}
\usepackage{tabularx}
\makesavenoteenv{tabular}
\makesavenoteenv{table}

\usepackage[printonlyused,withpage,nolist,nohyperlinks]{acronym}

\usepackage{xifthen}
\usepackage{xparse}
\usepackage{subdepth}
\usepackage{leftidx}
\usepackage{bm}
\usepackage{amsmath}

\newcommand{\bbm}{\begin{bmatrix}}
\newcommand{\ebm}{\end{bmatrix}}

\DeclareMathAlphabet{\mbf}{OT1}{ptm}{b}{n}
\newcommand{\mbs}[1]{{\bm{#1}}}

\newcommand{\mbsbar}[1]{{\overline{\boldsymbol{#1}}}}
\newcommand{\mbshat}[1]{{\hat{\boldsymbol{#1}}}}
\newcommand{\mbstilde}[1]{{\tilde{\boldsymbol{#1}}}}

\newcommand{\mbsdot}[1]{{\dot {\boldsymbol{#1}}}}

\newcommand{\mbfbar}[1]{{\overline{\mbf{#1}}}}
\newcommand{\mbfhat}[1]{{\hat{\mbf{#1}}}}
\newcommand{\mbftilde}[1]{{\tilde{\mbf{#1}}}}

\newcommand{\mbfdot}[1]{{\dot{\mbf{#1}}}}

\newcommand{\cframe}[1]{{\smash{\protect\underrightarrow{\mathcal{F}}_{#1}}}}

\DeclareMathAlphabet{\mathbfit}{OML}{cmm}{b}{it}
\newcommand{\homo}[1]{{\mathbfit{#1}}}

\newcommand{\mbfh}[1]{{\homo{#1}}}

\newcommand{\pos}[2]{\leftidx{_{#1}}{ \mbf r}{_{#2}}} 
\newcommand{\postilde}[2]{\leftidx{_{#1}}{\mbftilde r}{_{#2}}} 
\newcommand{\posh}[2]{\leftidx{_{#1}}{\mbfh r}{_{#2}}} 

\newcommand{\vel}[3]{\leftidx{_{#1}}{\mbf v}{\IfValueTF{#2}{_{#2#3\hspace{2pt}}}{}}} 
\newcommand{\veltilde}[3]{\leftidx{_{#1}}{\mbftilde v}{\IfValueTF{#2}{_{#2#3\hspace{2pt}}}{}}} 
\newcommand{\velbar}[3]{\leftidx{_{#1}}{\mbfbar v}{\IfValueTF{#2}{_{#2#3\hspace{2pt}}}{}}} 
\newcommand{\velhat}[3]{\leftidx{_{#1}}{\mbfhat v}{\IfValueTF{#2}{_{#2#3\hspace{2pt}}}{}}} 
\newcommand{\veldot}[3]{\leftidx{_{#1}}{\mbfdot v}{\IfValueTF{#2}{_{#2#3\hspace{2pt}}}{}}} 

\newcommand{\acc}[3]{\leftidx{_{#1}}{\mbf a}{\IfValueTF{#2}{_{#2#3\hspace{2pt}}}{}}} 
\newcommand{\acctilde}[3]{\leftidx{_{#1}}{\mbftilde a}{\IfValueTF{#2}{_{#2#3\hspace{2pt}}}{}}} 
\newcommand{\accbar}[3]{\leftidx{_{#1}}{\mbfbar a}{\IfValueTF{#2}{_{#2#3\hspace{2pt}}}{}}} 

\newcommand{\rotvel}[3]{\leftidx{_{#1}}{\mbs \omega}{\IfValueTF{#2}{_{#2#3\hspace{2pt}}}{}}} 
\newcommand{\rotveltilde}[3]{\leftidx{_{#1}}{\mbstilde \omega}{\IfValueTF{#2}{_{#2#3\hspace{2pt}}}{}}} 
\newcommand{\rotvelbar}[3]{\leftidx{_{#1}}{\mbsbar \omega}{\IfValueTF{#2}{_{#2#3\hspace{2pt}}}{}}} 
\newcommand{\rotvelhat}[3]{\leftidx{_{#1}}{\mbshat \omega}{\IfValueTF{#2}{_{#2#3\hspace{2pt}}}{}}} 
\newcommand{\rotveldot}[3]{\leftidx{_{#1}}{\mbsdot \omega}{\IfValueTF{#2}{_{#2#3\hspace{2pt}}}{}}} 

\newcommand{\T}[2]{\leftidx{}{\mbfh T}{_{#1#2\hspace{2pt}}}} 
\newcommand{\q}[2]{\leftidx{}{\mbf q}{_{#1#2\hspace{2pt}}}} 
\newcommand{\qtilde}[2]{\leftidx{}{\mbftilde q}{_{#1#2\hspace{2pt}}}} 
\newcommand{\qbar}[2]{\leftidx{}{\mbfbar q}{_{#1#2\hspace{2pt}}}} 

\newcommand{\Exp}[1]{\text{Exp}\left( #1 \right)}
\newcommand{\Log}[1]{\text{Log}\left( #1 \right)}

\title{\LARGE \bf
OKVIS2: Realtime Scalable Visual-Inertial SLAM with Loop Closure
\thanks{This work has been supported by the EPSRC grant Aerial ABM EP/N018494/1, EPSRC ORCA Partnership Resource Fund (PRF) SWIFT, Imperial College London, and the Technical University of Munich.}
}

\author{Stefan Leutenegger$^{1,2}$
\thanks{$^{1}$The author is with the Smart Robotics Lab, Technical University of Munich, Germany.
  {\tt\small stefan.leutenegger@tum.de}.}%
\thanks{$^{2}$The author is also with the Smart Robotics Lab, Imperial College London, UK.}%
}

\begin{document}

\maketitle
\thispagestyle{plain}
\pagestyle{plain}

\begin{abstract}
Robust and accurate state estimation remains a challenge in robotics, Augmented, and Virtual Reality (AR/VR), even as Visual-Inertial Simultaneous Localisation and Mapping (VI-SLAM) getting commoditised. Here, a full VI-SLAM system is introduced that particularly addresses challenges around long as well as repeated loop-closures. A series of experiments reveals that it achieves and in part outperforms what state-of-the-art open-source systems achieve. At the core of the algorithm sits the creation of pose-graph edges through marginalisation of common observations, which can fluidly be turned back into landmarks and observations upon loop-closure. The scheme contains a realtime estimator optimising a bounded-size factor graph consisting of observations, IMU pre-integral error terms, and pose-graph edges––and it allows for optimisation of larger loops re-using the same factor-graph asynchronously when needed.
\end{abstract}
\section{Introduction}
The ability to estimate states, such as pose and velocity, constitutes a fundamental capability for mobile robots enabling autonomy, or to power Augmented and Virtual Reality (AR/VR) applications. The combination of visual, spatial constraints with temporal ones provided by an Inertial Measurement Unit (IMU) has proven to be of particular interest providing high robustness and accuracy, therefore finding its way into various commercial systems in the AR/VR and mobile robotics space. 

Pushing the boundary of robustness and accuracy with open-source systems, while maintaining realtime operation, remains a much researched challenge. Consequently, the past few years have seen the emergence and release of several Visual-Inertial Odometry (VIO) and Simultaneous Localisation and Mapping (VI-SLAM) systems that exhibit impressive characteristics. Interestingly, state-of-the-art VIO and VI-SLAM systems such as BASALT \cite{usenko2019visual}, Kimera \cite{rosinol2020kimera}, VINS-Fusion \cite{qin2019b}, and ORB-SLAM3 \cite{campos2021orb} are all \emph{tightly-coupled} systems, i.e.\ jointly considering visual and inertial measurements with their internal states; and they further employ explicit \emph{sparse} data associations provided by keypoint detection and tracking and/or descriptor matching (indirect approach). 

Scalability to any environment, indoor and outdoor, as well as sensor suite and motion characteristic--especially when it comes to realtime handling of potentially frequent loop closures--still constitute a challenge, which this work aims to address: specifically, the scheme constructs pose-graph edges from joint observations that are used as part of the realtime estimator, jointly with visual reprojection error and pre-integrated inertial factors. This allows to increase the window of optimised states, therefore limiting detrimental effects of fixing older states. Upon loop closure, this incrementally constructed posegraph, together with inertial factors and observations, can be asynchronously optimised and, importantly, the posegraph edges can be seamlessly turned back into observations, thereby reviving old landmarks and reprojection errors. Furthermore, a light-weight segmentation Convolutional Neural Network (CNN) is run on keyframes asynchronously, on the CPU, to remove observations into dynamic regions. In a series of experiments on popular datasets, EuRoC \cite{Burri25012016} and \cite{schubert2018vidataset}, the competitive and in parts superior accuracy of OKVIS2 is demonstrated in comparison with state-of-the art approaches.

In summary, the following contributions are made:
\begin{itemize}
    \item A realtime capable multi-camera VI-SLAM system called OKVIS2 that supports place recognition and loop closure, which I aim to release as open-source software within the coming months.
    \item An algorithm constructing posegraph factors from marginalised observations, which are included into both the realtime tracking estimator to allow for an enlarged optimisation window, as well as in loop closure optimisation running asynchronously.
    \item A scheme of leveraging a light-weight semantic segmentation CNN for removal of dynamic objects (humans, clouds, \ldots) running on the CPU for maximum portability.
\end{itemize}

The remainder of this paper is organised as follows: in Section \ref{sec:related_work}, the most directly related work is reviewed, followed by a brief introduction to the notation used in Section \ref{sec:notations_and_definitions}, and a system overview in Section \ref{sec:overview}. Sections \ref{sec:estimator} and \ref{sec:frontend} then introduce the contributions in terms of estimator and the frontend, respectively. Experimental results are later presented in Section \ref{sec:results}.

\section{Related Work}
\label{sec:related_work}
We briefly overview the most directly relevant works on realtime vision-based and visual-inertial odometry and SLAM. 

\subsection{Visual Odometry and SLAM}
While very early vision-only SLAM systems used sparse observations in an Extended Kalman Filtering (EKF) scheme \cite{davison2007monoslam}, later systems quickly converged to employing non-linear least-squares optimisation thanks to superior computational efficiency and accuracy. PTAM \cite{klein2007parallel}, and later ORB-SLAM 1 \cite{mur2015orb} and 2 \cite{mur2017orb} all use some form of windowed Bundle Adjustment, the latter in combination with re-localisation and pose-graph optimisation upon loop closures. All of the above employ keypoint detection and tracking and/or descriptor matching, thus explicit correspondences.

A new paradigm was proposed with DTAM \cite{newcombe2011dtam}, which reconstructs dense depth frames and performs per-pixel direct photometric alignment for tracking, while maintaining realtime capabilities if implemented on a GPU. As such, DTAM offers a much more expressive map representation, however, tracking accuracy proved not quite competitive with sparse systems. As a compromise towards more computational efficiency, while maintaining direct photometric alignment, LSD-SLAM \cite{engel2015large} reconstructs depth in pixels of sufficient contrast only, and offers additionally loop closure detection and optimisation, therefore making for an efficient CPU-based full SLAM system. Finally, the concept of direct photometric alignment was also adopted for an otherwise sparse SLAM systems, SVO \cite{forster2014svo} and DSO \cite{engel2017direct}, the latter ultimately with stereo \cite{wang2017stereo} and loop closure \cite{gao2018ldso} extensions.

\subsection{Dense Depth-Camera-based SLAM}
The emergence of RGB-D (depth) cameras has fuelled the development of dense SLAM systems. KinectFusion \cite{newcombe2011kinectfusion} reconstructs a volumetric Truncated Signed Distance (TSDF) map and performs geometric alignment in tracking, by application of the Iterative Closest Point (ICP) algorithm for every pixel of the live frame. Various attempts to improve scalability in space followed, e.g.\ \cite{whelan2012kintinuous, vespa2019adaptive}. Alternative map representations have also been explored, e.g.\ surfels, which allowed for inclusion of loop closure in ElasticFusion \cite{whelan2016elasticfusion}. Common to all these systems is the alternation of tracking (optimisation of pose holding the map fixed) and mapping (holding the pose fixed and fusing new information into the map), which is the likely cause for inferior trajectory accuracy in large-scale scenarios. Note, however, that dense mapping backends may be combined with other, e.g.\ sparse visual (-inertial) SLAM systems.

\subsection{Visual-Inertial Odometry and SLAM}
Among the first attempts to tightly couple visual and inertial measurements, the seminal work \cite{mourikis2007multi} introduces the MSCKF, a filtering-based approach in which several past states are maintained and updated with observations under marginalisation of landmarks. Many later derived approaches improve accuracy and computational efficiency, making the method ideally suited in robotics, Augmented and Virtual Reality (AR/VR). An excellent open-source implementation is provided by OpenVINS \cite{geneva2020openvins}. ROVIO \cite{bloesch2017iterated} also employs a light-weight filtering scheme, but using a photometric update rather than indirect keypoint associations. Tight integration of loop closure and large-scale map management constitutes, however, an inherent challenge to schemes that employ marginalisation of old states and landmarks. An attempt of more loosely integrating large-scale sparse mapping and VIO is made by the popular framework Maplab \cite{schneider2018maplab}. 

Inspired by successes in vision-only works, a second stream of VIO and VI-SLAM systems employ nonlinear least-squares optimisation of typically reprojection error and integrated IMU measurements jointly, and in some form of a windowed approach to keep computational tractability. OKVIS \cite{leutenegger2015keyframe} achieves this by marginalisation and keeping a set of old keyframes held together with linearised error terms. ORB-SLAM3 \cite{campos2021orb}, extending \cite{mur2017orb}, conversely, fixes old states––a scheme that is simpler, more flexible to integrate with its loop closure optimisations, but also inherently not a conservative approximation, as it effectively ignores past estimation uncertainties. In an attempt to more rigorously assess the errors introduced by fixation, \cite{keivan2016asynchronous} analyses residual statistics and chooses the fixation adaptively leading to a shrinking and expanding state window that is optimised. This proposed work aims to alleviate the potential downsides of fixation by including a larger window of old keyframes into the optimisation problem, which is constrained by reletive pose factors, instead of solely computationally expensive observations.

It is worth noting that multi-sensor odometry and SLAM systems have much profited from open-source factor-graph optimisation libraries, most notably g2o \cite{grisetti2011g2o}, ceres \cite{agarwal2012ceres}, and GTSAM \cite{dellaert2012factor}. The latter also integrates iSAM 1 \cite{kaess2008isam} and 2 \cite{kaess2012isam2}, which allow to efficiently solve full-batch underlying factor graphs such as typical in VI-SLAM in an inherently incremental manner.

VINS-Fusion \cite{qin2019b} operates similarly to OKVIS, but includes loop closure by posegraph optimisation, where relative poses edges of consecutive-only keyframes are constructed from marginalisation of visual and inertial measurements. The system supports magnetometer and GPS measurements in addition to multiple cameras and an IMU. In contrast, this paper more rigorously constructs pose-graph edges from visual co-observations and uses these also as part of the realtime frame-by-frame tracking estimation.

Similarly, Kimera \cite{rosinol2020kimera} is a state-of-the-art sparse VI-SLAM system also supporting dense mesh mapping, that combines a VIO-frontend with a pose-graph optimisation backend used upon loop closure, both relying on GTSAM.

BASALT \cite{usenko2019visual} employs a similar idea as the one presented in this work. Non-linear factors connecting several keyframes are recovered from joint observations and inertial measurements. However, in contrast, in this work pre-integrated inertial factors are kept and only relative pose errors from observations common to two keyframes are computed.

\section{Notation and Definitions}
\label{sec:notations_and_definitions}
The VI SLAM problem consists of tracking a moving body with a mounted IMU and $N$ cameras relative to a static World coordinate frame $\cframe{W}$. The IMU coordinate frame is denoted as $\cframe{S}$ and the camera frames as $\cframe{C_i}$, $i=1 \ldots N$. 

Left-hand indices denote coordinate representation. Homogeneous position vectors (denoted in italic font) can be transformed with $\T{A}{B}$, meaning $\posh{A}{P} = \T{A}{B}\posh{B}{P}$. 

The following state representation is used:
\begin{equation}
\mbf{x} = [\pos{W}{S}^T, \q{W}{S}^T, \vel{W}{}{}^T, \mbf{b}_\mathrm{g}^T, \mbf{b}_\mathrm{a}^T]^T,
\end{equation}
where $\pos{W}{S}$ denotes the position of the origin of $\cframe{S}$ relative to $\cframe{W}$, $\q{W}{S}$ is the Hamiltonian Quaternion of orientation describing the attitude of $\cframe{S}$ relative to $\cframe{W}$, and $\vel{W}{}{}$ stands for the velocity of $\cframe{S}$ relative to $\cframe{W}$. Furthermore, the rate gyro biases $\mbf{b}_\mathrm{g}$ and accelerometer biases $\mbf{b}_\mathrm{a}$ are included. 
The state is estimated at every time step $k$ where frames from the camera(s) are obtained. 

For quaternions, we employ the multiplicative perturbation around a linearisation point $\qbar{A}{B}$
\begin{equation}
    \q{A}{B} = \Exp{{\delta \mbs{\alpha}_{AB}}} \otimes \qbar{A}{B},
\end{equation}
where $\Exp{.}$ stands for the quaternion group exponential that includes the transformation of a (small) rotation vector $\delta \mbs{\alpha}_{AB} \in \mathbb{R}^3$ into the Lie Algebra.

\section{System Overview}
\label{sec:overview}

\begin{figure}[t!]
\centering
\includegraphics[width=\columnwidth]{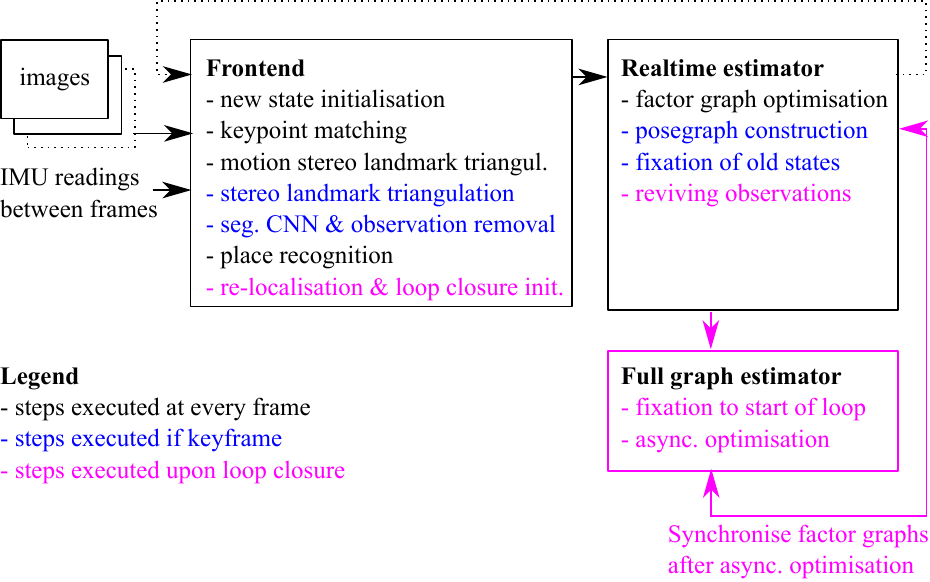}
\caption{Overview of OKVIS2: the frontend takes a multi-frame and associated IMU messages to initialise the new state, match keypoints, triangulate new landmarks, and to try place recognition, relocalisation, and loop closure. The realtime estimator will then optimise the underlying factor graph and take care of posegraph maintainance and fixation of old states. A full graph optimisation around detected loops is carried out asynchronously and later synchronised.}
\label{fig:overview}
\end{figure}

The VI SLAM system is split into \emph{frontend} and \emph{realtime estimator} that process images and IMU messages synchronously whenever a new (multi-) frame arrives. To deal with loop closures, a \emph{full factor graph} loop optimisation is executed asynchronously. As shown in Fig.~\ref{fig:overview}, the frontend deals with state initialisation, keypoint matching, stereo triangulation (of successive frames and from stereo images of the same multi-frame), running the segmentation CNN, as well as place recognition and, if the latter was successful, re-localisation and loop closure initialisation. The realtime estimator will then optimise the respective factor graph, and is also responsible for the creation of posegraph edges by marginalising old observations, as well as for fixation of old states. Upon loop closure, it further turns posegraph edges back into observations, and then triggers the optimisation of the full factor graph around a loop, which runs asynchronously, and which will be synchronised with the realtime factor graph upon completion.

\section{Visual-Inertial Estimator}
\label{sec:estimator}
In the following, the different components of the VI Estimator are outlined. 
Parts of the frontend, as well as visual and inertial error terms are largely adopted from OKVIS~\cite{leutenegger2015keyframe}.
The non-linear least squares costs as described below are minimised using Google's Ceres Solver \cite{ceres-solver}. 

The estimator will be minimising visual, inertial, and relative pose errors (pose graph edges), which are described separately in what follows.

\subsection{Reprojection Error}
We use the standard reprojection error $\mbf e^{i,j,k}_\mathrm{r}$ of the $j$-th landmark ${_{W}}\mbf{l}^j$ into the $i$-th camera image at time step $k$: 
\begin{equation}
\label{e:reprojection_error}
\mbf e^{i,j,k}_\mathrm{r} = \mbftilde z^{i,j,k}_\mbf{r} - \mbf{h}(\T{S}{C_i}^{-1}\T{S^k}{W} {_{W}}\mbfh{l}^j),
\end{equation}
with the keypoint detection $\mbftilde z^{i,j,k}_\mbf{r}$ and $\mbf{h}(.)$ denoting the camera projection, where in this work pinhole projection is used, optionally with distortion (radial-tangential or equidistant). Jacobians are omitted here, as they are analogous to \cite{leutenegger2015keyframe}.

\subsection{IMU Error}
The IMU error $\mbf e^{k_0}_\mathrm{s}$ between time instance $k$ and $n$ is used:
\begin{equation}
\mbf e^{k}_\mathrm{s} = \mbfhat{x}^{n}(\mbf{x}^{k}, \mbftilde{z}_\mathrm{s}^{k,n})) \boxminus \mbf x^{n} \in \mathbb{R}^{15},
\end{equation}
with $\mbfhat{x}^{n}(\mbf{x}^{k})$ denoting the predicted state at step $n$ based on the estimate $\mbf x^{k}$ and the IMU measurements (rate gyro and accelerometer readings) $\mbftilde{z}_\mathrm{s}^{k,n}$ from in-between these frames. 
The $\boxminus$ operator becomes the regular subtraction for all states but quaternions, where it is defined as 
\begin{equation}
\q{}{}\boxminus\q{}{}' = \Log{\q{}{}\otimes\q{}{}'^{-1}},
\end{equation}
with the $\Log{.}$ denoting the quaternion group logarithm that includes the transformation from the Lie Algebra to a rotation vector.

A formulation of this error term using a pre-integration scheme adopted from \cite{forster2016manifold} is used, rendering its evaluation tractable for any number of IMU samples used. 

\subsection{Relative Pose Error}
Furthermore, relative pose errors $\mbf e^{k,r}_\mathrm{p}$ between time steps $r$ and $k$ are used:
\begin{equation}
\label{e:posegraph_error}
\mbf e^{r,c}_\mathrm{p} = \mbf{e}_{\mathrm{p},0}^{r,c} +
\begin{bmatrix}
\pos{S^r}{S^c}-\postilde{S^r}{S^c}\\
\q{S^r}{S^c}\boxminus\qtilde{S^r}{S^c}
\end{bmatrix},
\end{equation}
where $\mbf{e}_0^{r,c}$ denotes a constant (to be explained below), and $\postilde{S^k}{S^r}$, $\qtilde{S^c}{S^r}$ stand for nominal relative position and orientation, respectively (expressed in the reference IMU coordinate frame $\cframe{S^r}$).

\subsection{Realtime Estimation Problem}
The realtime estimator running at least at camera frame rate will minimise the following non-linear least squares cost
\begin{equation}
\label{e:cost}
\begin{split}
c(\mbf x) &= \frac{1}{2} \displaystyle \sum_i \sum_{k \in \mathcal{K}} \sum_{j \in \mathcal{J}(i,k)} \rho\left({\mbf{e}^{i,j,k}_\mathrm{r}}^T \mbf{W}_\mathrm{r}  \mbf{e}^{i,j,k}_\mathrm{r}\right) \\
&+ \frac{1}{2} \sum_{k \in \mathcal{P}\bigcup\mathcal{K}\backslash f} {\mbf{e}^{k}_\mathrm{s}}^T \mbf{W}_\mathrm{s}^k  \mbf{e}^{k}_\mathrm{s}
+ \frac{1}{2} \sum_{r \in \mathcal{P}} \sum_{c \in \mathcal{C}(r)} {\mbf{e}^{r,c}_\mathrm{p}}^T \mbf{W}_\mathrm{p}^{r,c} \mbf{e}^{r,c}_\mathrm{p}.
\end{split}
\end{equation}
Here, $\mbf{W}_\mathrm{r}$ stands for the visual weight as the inverse covariance of the reprojection error, and $\mbf{W}_\mathrm{s}^{k}$ for the IMU error weight. Further, the Cauchy robustifier $\rho(.)$ is used on observations. The set $\mathcal{K}$ denotes all poses with observations of the respectively visible landmarks in the set $\mathcal{J}(i,k)$. $\mathcal{K}$ contains the $T$ most recent frames as well as $M$ keyframes in the past; frame $f$ denotes the current frame. The set $\mathcal{P}$ contains the posegraph frames, i.e.\ those linked together with relative pose errors obtained from original co-observations. The weight $\mbf{W}_\mathrm{p}^{r,c}$ is computed from marginalisation of old observations, $\mathcal{K}$. Note that $\mathcal{P}$ reaches substantially further into the past than the frames with observations. The set $\mathcal{C}(r) \subset \mathcal{P}$ describes all the posegraph frames connected to frame $r$ with a posegraph edge. IMU errors are considered for all (key-) frames $k,n$ in succession. The IMU error weight $\mbf{W}_\mathrm{s}^k$ is obtained from linearised error propagation as part of the IMU error (pre-) integration. 

A new frame is selected as a keyframe based on a co-visibility criterion with currently active keyframes, further detailed in Section \ref{sec:keyframing}.

If the oldest frame in the $T$ most recent frames is not a keyframe, the respective states are simply removed and IMU measurements are appended to the previous IMU error using the pre-integration scheme up to the next frame. If, however, the oldest frame in the $T$ most recent frames is a keyframe, we apply the posegraph creation scheme detailed below in Section \ref{sec:posegraph_creation}.

\begin{figure}[t!]
\centering
\subfloat[In the beginning.]{
\label{fig:leutenegger_diagram_1}
\includegraphics[height=1.55cm]{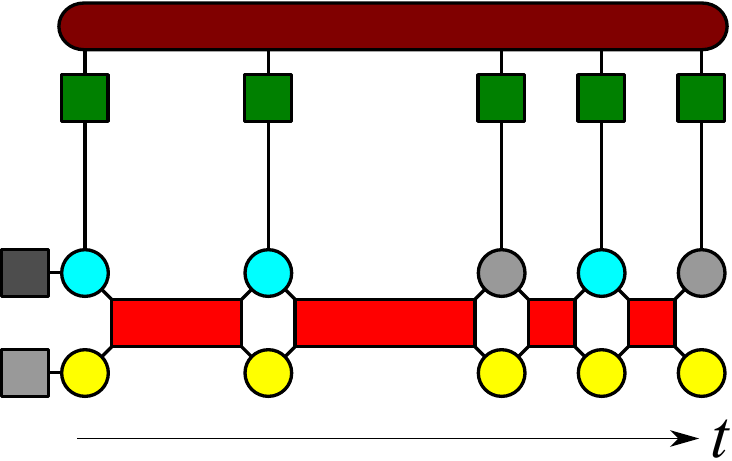}}\,
\subfloat[Later: posegraph creation.]{
\label{fig:leutenegger_diagram_2}
\includegraphics[height=1.55cm]{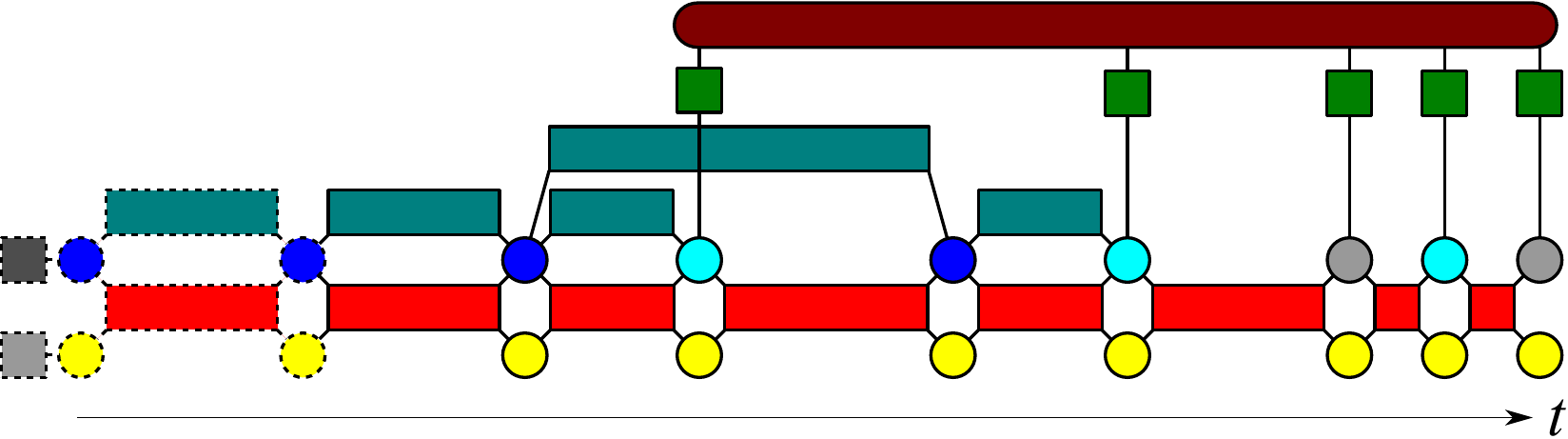}}\\
\subfloat[Loop closure.]{
\label{fig:leutenegger_diagram_3}
\includegraphics[height=1.55cm]{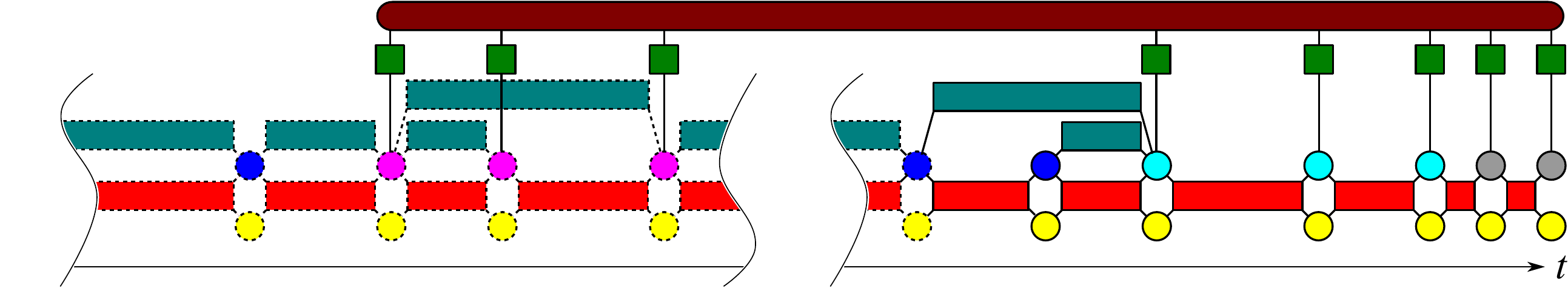}}\\
\vspace{0.2cm}
\includegraphics[height=2.7cm]{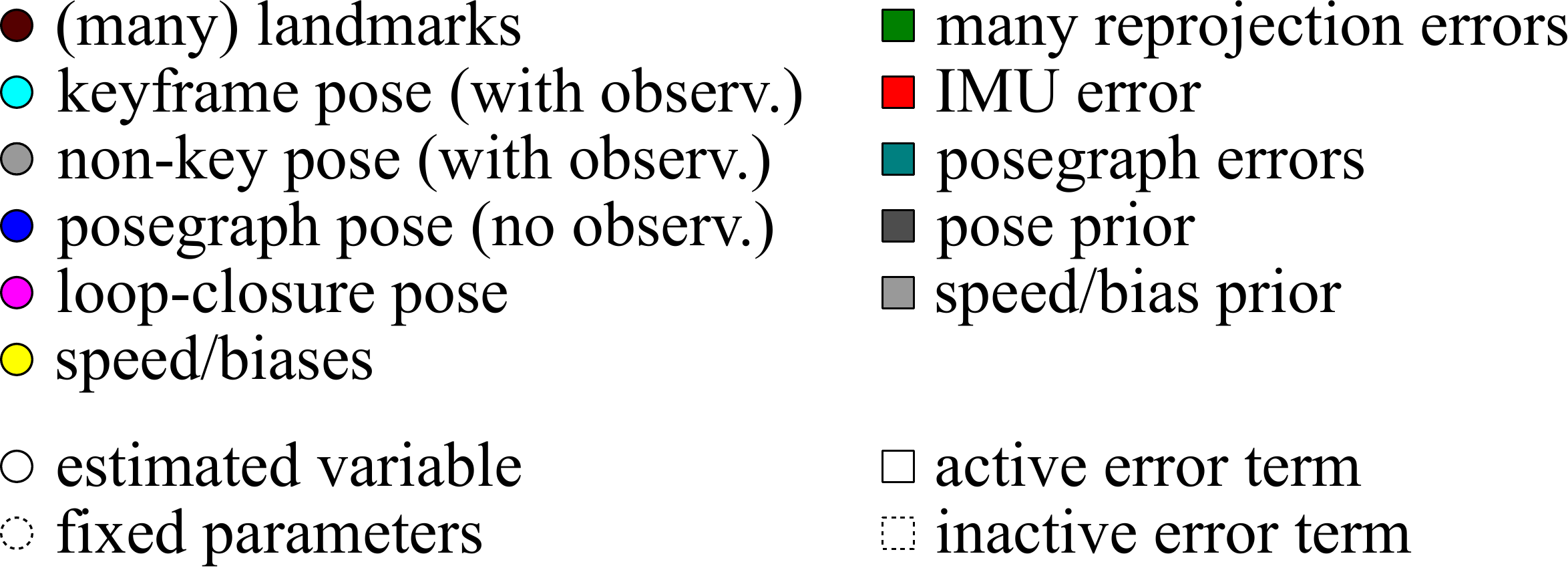}\\  
\caption{Initially a full batch VI factor graph (Fig.~\ref{fig:leutenegger_diagram_1}) is created and optimised. Later (Fig.~\ref{fig:leutenegger_diagram_2}), frames with least overlap with the live frame and current keyframe are turned into posegraph poses by construction of relative pose errors under marginalisation of common observations; also, old poses and speed/bias variables are fixed (dashed lines) to keep the problem realtime capable. When a loop-closure occurs (Fig.~\ref{fig:leutenegger_diagram_3}), respective observations and landmarks are re-activated.}
\label{fig:leutenegger_diagram}
\end{figure}

Fig.~\ref{fig:leutenegger_diagram} visually overviews the construction of the underlying factor graph as time progresses, which is further detailed below.

Note that while not further explained here, we may optionally leave the camera extrinsics $\T{S}{C_i}$ as variables to be optimised, i.e.\ performing online calibration.

\subsubsection{Posegraph Creation}
\label{sec:posegraph_creation}
To keep the problem complexity bounded, whenever the number of keyframes $|\mathcal{K}|$ exceeds a bound $K$, the frames exhibiting least co-visibility with either the current frame or current keyframe are moved from $\mathcal{K}$ to $\mathcal{P}$ by marginalisation of common observations under insertion of relative pose error terms of the form (\ref{e:posegraph_error}). The current keyframe is determined as the frame with most visual overlap with the current (most recent) frame (see also \ref{sec:frontend}).

One important exception is applied: the oldest keyframe is kept for as long as there are common observations with either the current frame or current keyframe, in order to keep some very long-term co-visibilities helping to maintain an accurate directional estimate.

\begin{figure}[t!]
\centering
\subfloat[Keyframe pose $r$ selected for conversion into posegraph node.]{
\label{fig:edge_creation_1}
\includegraphics[height=2.1cm]{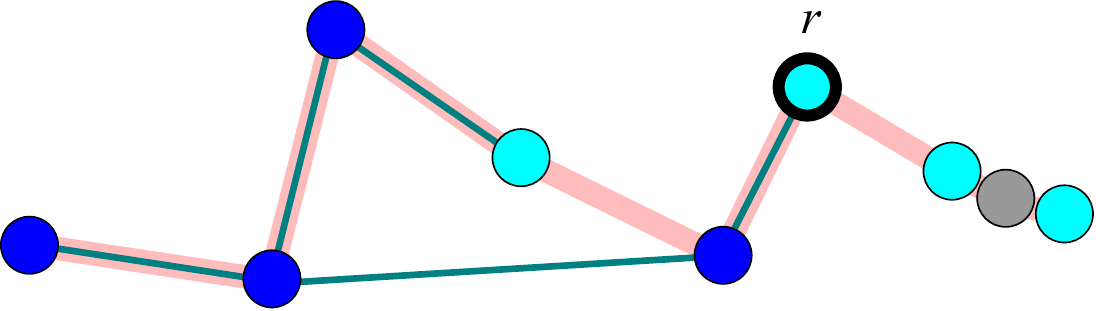}}\,
\subfloat[Computation of the Maximum Spanning Tree (MST).]{
\label{fig:edge_creation_2}
\includegraphics[height=2.1cm]{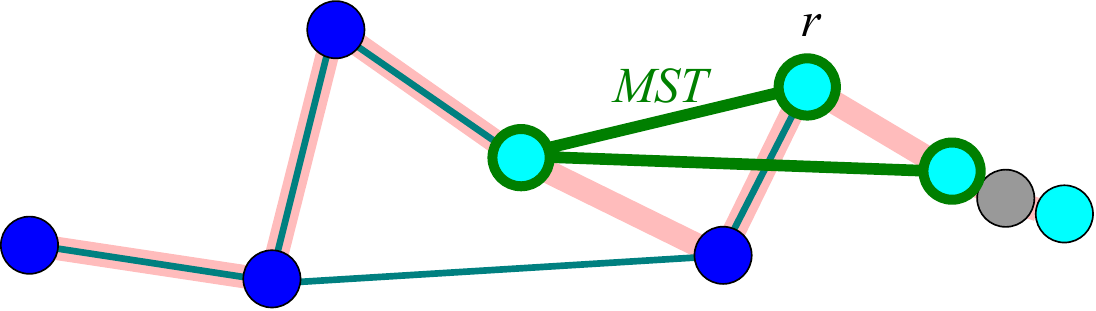}}\\
\subfloat[Creation of pose graph edges (here: just one).]{
\label{fig:edge_creation_3}
\includegraphics[height=2.1cm]{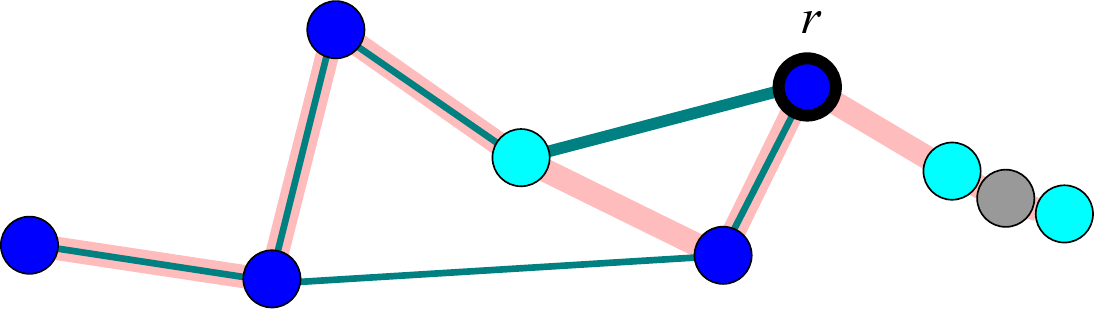}}\\
\vspace{0.2cm}
\includegraphics[height=1.1cm]{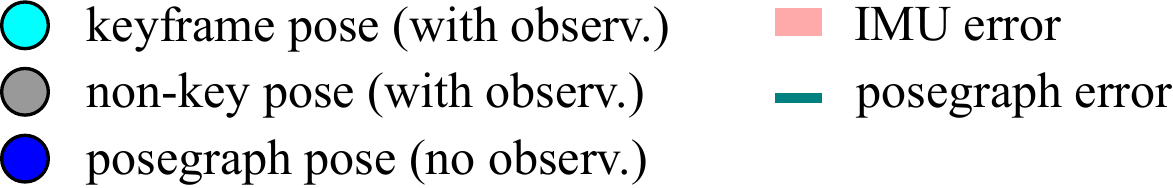}\\  
\caption{Posegraph edge creation scheme: when keyframe $r$ is selected for conversion into a posegraph node, because of least overlap with the current (key-)frame (Fig.~\ref{fig:edge_creation_1}), a Maximum Spanning Tree (MST) is computed \ref{fig:edge_creation_2}) from the poses in the set $\mathcal{T}$ (nodes with posegraph edges and node with most visual overlap), after which all MST edges connected to $r$ are created (Fig.~\ref{fig:edge_creation_3}).}
\label{fig:edge_creation}
\end{figure}

To create only a tractable subset of pose graph edges, the following heuristic is applied, which is visually explained in Fig.~\ref{fig:edge_creation}. 
\begin{itemize}
    \item From the poses with observations (i.e.\ in $\mathcal{K}$), those which already have posegraph edges connected are collected and inserted into a set $\mathcal{T}$.
    \item Let the pose index $r$ denote the states to be converted into a posegraph node. $r$ is inserted into $\mathcal{T}$.
    \item The frame with maximum visual overlap with $r$ is inserted into $\mathcal{T}$.
    \item A Maximum Spanning Tree (MST) is computed on $\mathcal{T}$ using the number of co-observations between two poses as the criterion.
    \item Now, all the MST edges connecting to the pose $r$ are created with the scheme described below, which uses all joint observations into the two frames $r$ and $c$.
\end{itemize}

Note that this method will ultimately duplicate observations and landmarks, if observed in more than the two frames $r$ and $c$, which introduces an inconsistency by design. However, the argument is made that this approximation will still produce superior results than simply creating relative pose errors with identity weight matrices, which is the de-facto standard in visual(-inertial) SLAM systems\ldots

Specifically, we first transform the landmark into the reference coordinates $\cframe{S^r}$, i.e.\ the IMU coordinate frame at index $r$. And we temporarily view as the state solely $\mbf{p} = [\pos{S_r}{S_c}, \q{S_r}{S_c}]$. Therefore, the reprojection errors into reference frame index $r$ and the other frame with index $c$, respectively, become 
\begin{eqnarray}
    \mbf{e}_\mathrm{r}^{i,j,r} &= &\mbftilde z^{i,j,r}_\mbf{r} - \mbf{h}(\T{S}{C_i}^{-1}{_{S^r}}\mbfh{l}^j),\\
    \mbf{e}_\mathrm{r}^{i,j,c} &= &\mbftilde z^{i,j,c}_\mbf{r} - \mbf{h}(\T{S}{C_i}^{-1}\T{S^c}{S^r} {_{S^r}}\mbfh{l}^j).
\end{eqnarray}.

Adopting landmark marginalisation to all joint observations, we consider the respective joint cost 
\begin{equation}
\label{e:posegraph_cost}
    c_\mathrm{p}^{r,c} = \frac{1}{2}\displaystyle \sum_{i=1}^N 
    \sum_{\substack{j\in\\\mathcal{J}(i,r)}}{\mbf{e}_\mathrm{r}^{i,j,r}}^T \mathbf{W}_\mathrm{r} \mbf{e}_\mathrm{r}^{i,j,r} + 
    \sum_{\substack{j\in\\\mathcal{J}(i,c)}}{\mbf{e}_\mathrm{r}^{i,j,c}}^T \mathbf{W}_\mathrm{r} \mbf{e}_\mathrm{r}^{i,j,c}
    ,
\end{equation}
with the sets $\mathcal{J}(i,r)$, $\mathcal{J}(i,c)$ denoting all landmarks that are visible at time steps $r$ and $c$ that have a measurement in the respective image $i$, and that have a low reprojection error to start with. Now we can approximate this cost as 
\begin{eqnarray}
\label{e:posegraph_cost_approx}
    c_\mathrm{p}^{r,c} &\approx &\frac{1}{2} {\mbf{e}_\mathrm{p}^{r,c}}^T\mbf{W}_\mathrm{p}^{r,c}\mbf{e}_\mathrm{p}^{r,c},\\
\label{e:posegraph_error2}
    \mbf{e}_\mathrm{p}^{r,c} &= &\mbf{e}_{\mathrm{p},0}^{r,c} +
\begin{bmatrix}
\pos{S^r}{S^c}-\postilde{S^r}{S^c}\\
\q{S^r}{S^c}\boxminus\qtilde{S^r}{S^c}
\end{bmatrix},
\end{eqnarray}
where $\postilde{S^r}{S^c}$, $\qtilde{S^r}{S^c}$ are the relative position and orientation, respectively, as a selected linearisation point at the time of pose graph edge creation. 
While the derivation of the Jacobians is omitted for brevity, note that for the important case of $\q{S^r}{S^c}$ approaching $\qtilde{S^r}{S^c}$, the Jacobian becomes Identity.

The remaining terms in (\ref{e:posegraph_cost_approx}) are computed applying the Schur complement operation on the Gauss-Newton system underlying the cost in (\ref{e:posegraph_cost}) for marginalisation of landmarks as follows. First, we construct the Gauss-Newton system that will assume the following structure
\begin{equation}
    \begin{bmatrix}
    \mbf{H}_{\mathrm{p},\mathrm{p}} 
    & \ldots & \mbf{H}_{\mathrm{p},j} & \ldots\\
    \vdots &\ddots &\mbf{0} &\mbf{0}\\
    \mbf{H}_{\mathrm{p},j}^T &\mbf{0} &\mbf{H}_{j,j} &\mbf{0}\\
    \vdots &\mbf{0} &\mbf{0} &\ddots\\
    \end{bmatrix} 
    \begin{bmatrix}
    \delta \mbf{p}\\
    \vdots\\
    \delta \mbf{l}_j \\
    \vdots
    \end{bmatrix}
    = 
    \begin{bmatrix}
    \mbf{b}_{\mathrm{p}} \\
    \vdots\\
    \mbf{b}_{j}\\
    \vdots
    \end{bmatrix}.
\end{equation}
Please note the variable ordering of the pose ($\mbf{p}$) first, followed by all the landmarks (${_{S^r}}\mbf{l}^j$). While the derivation of the necessary Jacobians is omitted here, they directly follow from the reprojection error term (\ref{e:reprojection_error}) with the notable property that landmarks observed in the reference frame need not be transformed, as they already are expressed in the reference frame, therefore the respective Jacobians become zero. 

Now, landmarks are marginalised out using the Schur complement:
\begin{eqnarray}
\mbf{H}^* &= &\mbf{H}_{\mathrm{p},\mathrm{p}} 
- \sum_j \mbf{H}_{\mathrm{p},j} \mbf{H}_{jj}^+ \mbf{H}_{\mathrm{p}, j}^T,\\
\mbf{b}^* &= &\mbf{b}_\mathrm{p} 
- \sum_j \mbf{H}_{\mathrm{p},j} \mbf{H}_{jj}^+ \mbf{b}_j,
\end{eqnarray}
which yields the reduced Gauss-Newton system
\begin{equation}
    \mbf{H}^* \delta \mbf{p} = \mbf{b}^*,
\end{equation}
which is now linearised around the current pose $\mbftilde{p}$, therefore 
\begin{equation}
\label{e:reduced_GN}
    \mbf{H}^* \delta \mbf{p} = \mbf{b}^* - \mbf{H}^* (\mbf{p}\boxminus\mbftilde{p})
\end{equation}
The supposedly equivalent Gauss-Newton system of a respective pose graph error term takes the form
\begin{equation}
    {\mbf{E}_\mathrm{p}^{r,c}}^T \mbf{W}_\mathrm{p}^{r,c}\mbf{E}_\mathrm{p}^{r,c} \delta \mbf{p} = -{\mbf{E}_\mathrm{p}^{r,c}}^T \mbf{W}_\mathrm{p}^{r,c}(\mbf{e}_\mathrm{p,0}^{r,c}+\mbf{p}\boxminus\mbfbar{p}),
\end{equation}
Where $\mbf{E}_\mathrm{p}^{r,c}$ denotes the Jacobian. For this to be equivalent to (\ref{e:reduced_GN}) at the linearisation point (i.e.\ at $\mbf{p}=\mbftilde{p}$ where $\mbf{E}_\mathrm{p}^{r,c}=\mbf{I}_6$), we now obtain
\begin{eqnarray}
\mbf{W}_\mathrm{p}^{r,c} = \mbf{H}^*,\\
\mbf{e}_{\mathrm{p},0}^{r,c} = -{\mbf{H}^*}^+ \mbf{b}^*.
\end{eqnarray}
Since $\mbf{H}^*$, $\mbf{b}^*$, and $\mbftilde{p}$ are constants, the Jacobians w.r.t.\ the poses at steps $r$ and $c$ are fairly straightforward to determine after substituting $\T{S^r}{S^c} = \T{W}{S^r}^{-1}\T{W}{S^c}$.

\subsubsection{Fixation of Old States}
Furthermore, in order to keep the number of poses being estimated limited, only the $A$ most recent states are kept variable. The number $A$ is determined as the maximum of a constant parameter $A_\mathrm{min}$ and $A_{\Delta T}$, the latter standing for the number of states covering the time interval $\Delta T$ into the past from the current frame.

\subsection{Place Recognition, Relocalisation, and Loop Closure}
Whenever a query of the current frame to the DBoW2~\cite{GalvezTRO12} database returns a match, say to pose index $l$, and geometric verification using 3D-2D RANSAC passes, the currently active window of states and landmarks is transformed to be re-aligned to the matched pose $l$. Then, the pose graph edges connecting $l$ will be ``revived'' and turned back into landmarks and observations; and observations with the current matching frame will also be created. This may also trigger merging of landmarks, if already existing new landmarks are matched to old landmarks in $l$. This scheme will produce a set of loop-closure frames, $\mathcal{L}$, that now has observations again; but it will in general be part of the fixed states of the realtime estimation graph, as these loop closure frames tend to lie arbitrarily far in the past. This concludes the relocalisation.

To optimise the loop inconsistency, the following scheme is applied: first, the error is distributed around the loop using rotation averaging followed by position inconsistency distribution, all in equal parts to the (keyframe) poses around the loop. Then, a background optimisation process of the factor graph is triggered. Hereby, a copy of the realtime estimator graph is used, however, with different fixation of states (i.e.\ the states inside the closed loop remain variable). Note how all the IMU error terms are also considered in this loop closure optimisation, adding further constraints, most notably regarding consistency of the orientation relative to gravity. After the optimisation has finished, a synchronisation process imports the optimised states and landmarks into the realtime estimator, and re-aligns new states and landmarks created in the meantime. During this optimisation, relocalisation attempts are suspended. Upon completion, the number of loopclosure frames $|\mathcal{L}|$ is limited to a constant $L$ applying the same pose graph edge creation scheme as \ref{sec:posegraph_creation}.

\section{Frontend Overview}
\label{sec:frontend}
In the following, the remaining elements of the frontend are briefly overviewed.

\subsection{Matching and initialisation}
The visual frontend extracts BRISK 2 \cite{leutenegger2011brisk, leutenegger2014unmanned} keypoints and descriptors in every image of a multi-frame, and matches them to the 3D landmarks already in the map; hereby both descriptor distance and-- conservatively--reprojected image distance are considered. 
New 3D landmarks are then initialised both from stereo triangulation within the $N$ images of a keyframe, as well as from triangulation between live frame and any of the keyframes.

\subsection{Keyframe selection}
\label{sec:keyframing}
The decision of whether a new frame is considered a keyframe is taken depending on the fraction of matched landmarks in the live frame $o_\mathrm{l}$, as well as the fraction of current matches visible in any of the existing keyframes (with index $k\in\mathcal{K}$), $o_k$. Specifically, these overlap fractions are determined by the number of matched keypoint areas divided by the total of all keypoint areas in the respective (multi-image) frames; whereby the respective keypoint areas are obtained from the respective union of filled circles of radius $r_\mathrm{kpt}$ around the keypoints. If the overlap 
\begin{equation}
    o = \min(o_\mathrm{l}, \max_{k \in \mathcal{K}}o_k),
\end{equation} falls below a threshold $t_o$, we accept the (multi-)frame as a new keyframe.

\subsection{Segmentation CNN}
\begin{figure}[!htb]
\centering
\includegraphics[height=4.2cm]{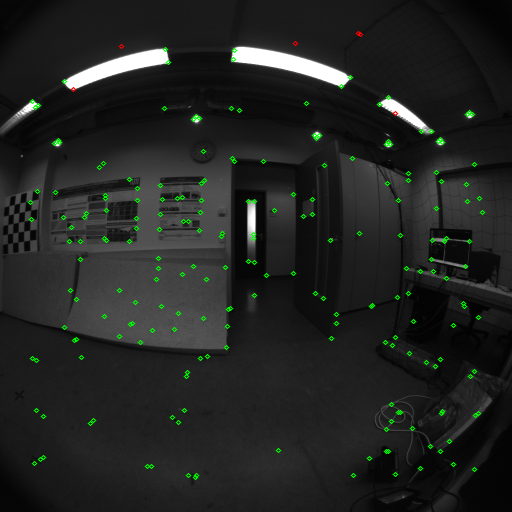}
\includegraphics[height=4.2cm]{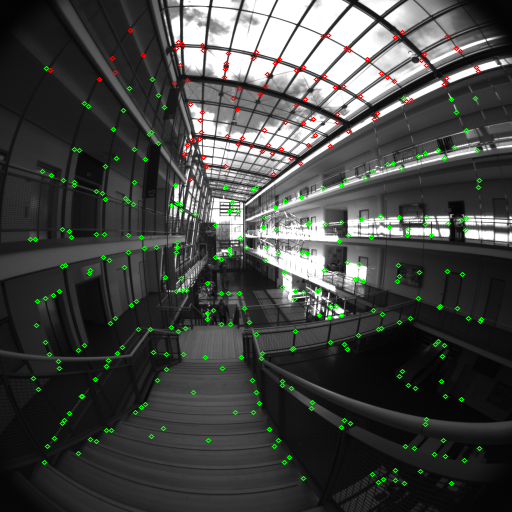}\\\vspace{1.3mm}
\includegraphics[height=4.2cm]{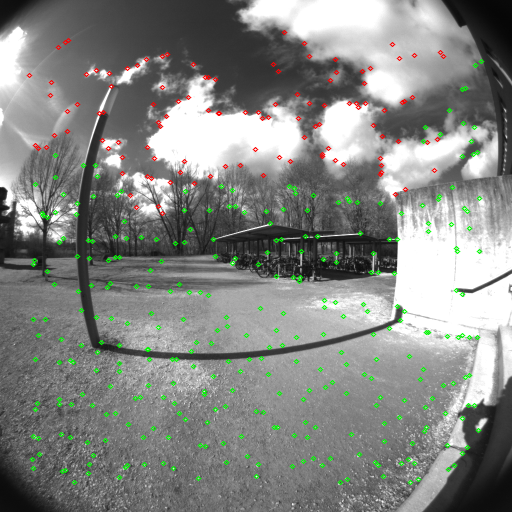}
\includegraphics[height=4.2cm]{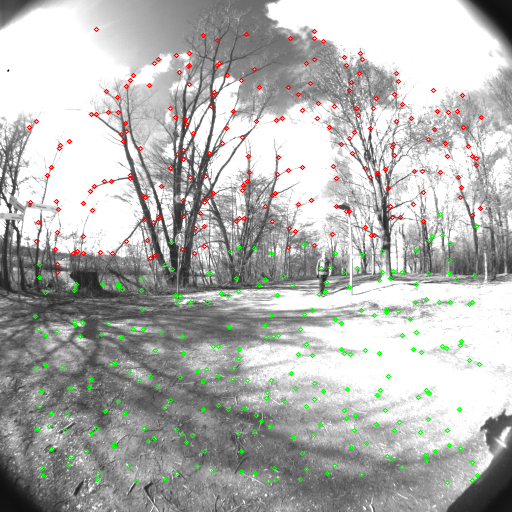}
\caption{TUM-VI \texttt{outdoors1} example sky classifications (red) vs.\ used keypoints: indoors (top), and outdoors (bottom). Notice how some erroneous sky classifications occur also indoors. Further it can be seen that most of the sky keypoints are identified correctly outdoors, but there are some false negatives (on the rightmost cloud) and false positives (close to the horizon, especially on the leaf-less tree branches).}
\label{fig:classified}
\end{figure}

The semantic segmentation network Fast-SCNN \cite{poudel2019fast} is run on all images of a keyframe. The network was fine-tuned on the Cityscapes dataset \cite{Cordts2016Cityscapes} for 80 epochs, where images were first converted to grayscale, in order to best perform with gray images used as input to OKVIS2. For maximum portability and flexibility, inference is carried out as an asynchronous background process on the CPU, which is tractable, due to the efficient network as well as the fact that only keyframes are processed. Once completed, matches into images regions that are almost guaranteed to be dynamically changing are simply removed (currently only the sky, but people, animals, etc.\ could be considered with appropriate training). 
As will be shown in the experiments, this scheme significantly improves accuracy in presence of slow-moving scene content, particularly clouds, where observations aren't already automatically discarded via the Cauchy robustification. Please see Fig.~\ref{fig:classified} for example classifications.

\section{Experimental Results}
\label{sec:results}
Results are provided for two commonly used datasets in VI Odometry/SLAM evaluation: the ETHZ-ASL EuRoC Dataset \cite{Burri:JFR2018} and the TUM VI Benchmark \cite{schubert2018vidataset},
all providing ground truth trajectories, but featuring different characteristics in cameras, motion, (visual) environment and spatial extent of the trajectories.

Two quantities are evaluated that the author strongly believes should be treated separately. 
\begin{itemize}
 \item The first one is the accuracy in terms of \emph{odometry drift}, where loop-closures are disabled. The same methodology as in \cite{leutenegger2015keyframe} is used, where statistics of relative motion error (position/orientation) are aggregated for in buckets of distance travelled. For completeness, relative position/orientation errors with loop-closure enabled are also reported, but note that the results are consequently highly dependent on the motion, i.e.\ how frequently loop-closures occur.
 \item The second one is the \emph{Absolute Trajectory Error (ATE)} where loop closure is taken into account, and where we simply align the estimated and ground truth trajectories in terms of position and yaw angle (roll and pitch are globally observable and are thus not aligned) and report statistics of the position differences. Here, it is important to note that we can either evaluate the estimates that use all measurements up to the respective timestamp, i.e.\ in a \emph{causal} manner, or the final fully loop-closed trajectory where all measurements are taken into account for all estimates, i.e.\ as an \emph{non-causal} evaluation. Both are valid methods, but it should be stated clearly; and note that in the former case, the drift (as already assessed in the odometry evaluation) may dominate the ATE error statistics as the trajectory will contain the characteristic ``jumps'' when correcting for loop-closures.
\end{itemize}

The same estimator parameters were used throughout, namely the $T=3$ most recent frames are kept, $K=5$ keyframes, and $L=5$ loop-closure frames. Furthermore, at least $A_\mathrm{min}=12$ pose-graph frames are optimised, where all frames no older than $\Delta T= 2$ sec are also always optimised. Frontend and IMU parameters are dataset-specific, and respective configuration files will be released together with the code.

\subsection{ETH-ASL EuRoC Dataset}
\begin{figure}[!htb]
    \centering
    \includegraphics[width=\columnwidth, trim={1.0cm 0.2cm 1.2cm 0.5cm},clip]{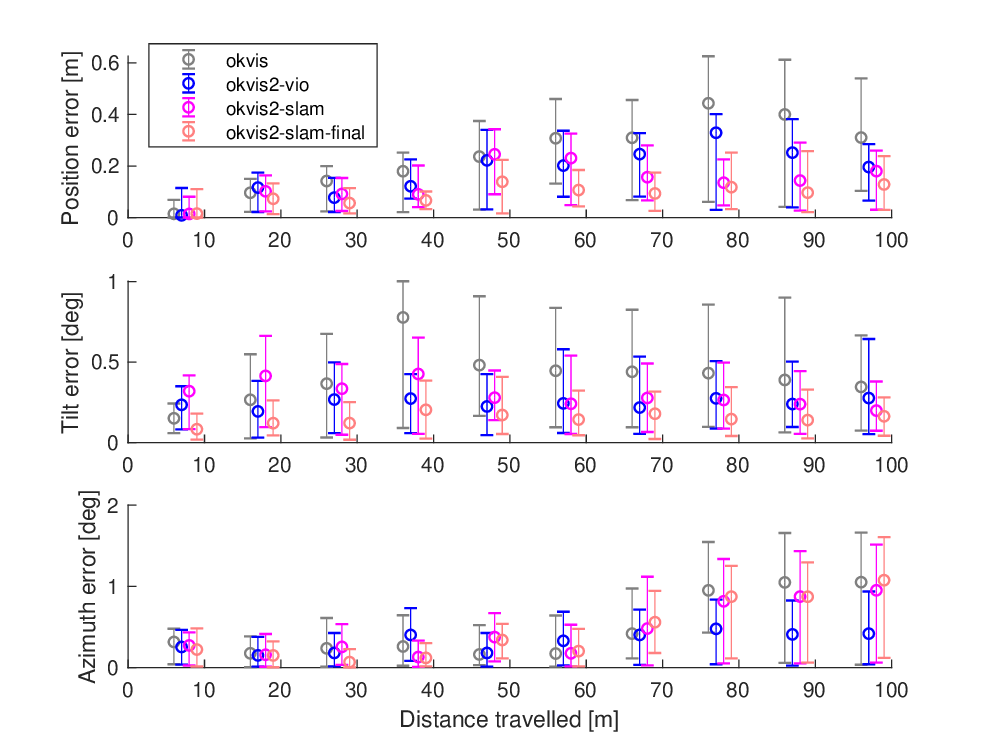}
    \caption{Relative trajectory error statistics for different sub-trajectory lengths (y-axis) of the exemplary EuRoC \texttt{MH\_05\_difficult} sequence.}
    \label{fig:euroc_relative}
\end{figure}

Relative error statistics for the \texttt{MH\_05\_difficult} sequence of the on the EuRoC Dataset \cite{Burri25012016} are shown in Fig.~\ref{fig:euroc_relative} as an illustrative example of how OKVIS2 in VIO mode, \texttt{okvis2-vio}, (i.e.\ with disabled loop closures) improves over original OKVIS \cite{leutenegger2015keyframe}, \texttt{okvis}, and how loop closures will reduce long-term drift. 

What can also be easily seen is how the smoothness of the non-causal evaluation of the final optimised trajectory \texttt{okvis2-slam-final} leads to substantially higher accuracy over the causal version \texttt{okvis2-slam}.

In Table \ref{t:euroc}, Absolute Trajectory Error (ATE) results are presented, also in comparison to state-of-the-art competitors from both the VIO and VI-SLAM world.
\begin{table}[!htb]
\setlength\tabcolsep{4.5pt} 
\begin{tabularx}{\columnwidth}{l|rrrrrr|rr}
& \rotatebox[origin=lb]{90}{\textbf{BASALT} \cite{usenko2019visual}$^1$}
& \rotatebox[origin=lb]{90}{\textbf{Kimera} \cite{rosinol2020kimera}$^1$} 
& \rotatebox[origin=lb]{90}{\textbf{OKVIS} \cite{leutenegger2015keyframe}$^2$
} 
& \rotatebox[origin=lb]{90}{\textbf{OKVIS2 VIO (ours)}} 
& \rotatebox[origin=lb]{90}{\textbf{VINSFusion} \cite{qin2019b}$^1$} 
& \rotatebox[origin=lb]{90}{\textbf{OKVIS2 causal (ours)}} 
& \rotatebox[origin=lb]{90}{\textbf{ORB-SLAM3} \cite{campos2021orb}$^1$}
& \rotatebox[origin=lb]{90}{\textbf{OKVIS2 (ours)}} \\
\midrule
& \multicolumn{6}{c|}{Causal evaluation} & \multicolumn{2}{c}{Non-causal} \\
\midrule
& \multicolumn{4}{c|}{VIO} & \multicolumn{4}{c}{VI-SLAM} \\
\midrule
\textbf{\texttt{MH\_01}} & 0.080  & 0.080  & 0.079  & 0.057  & 0.166   & 0.044  & 0.036 & \textbf{0.027} \\
\textbf{\texttt{MH\_02}} & 0.060  & 0.090  & 0.044  & 0.044  & 0.152   & 0.036  & 0.033 & \textbf{0.023} \\
\textbf{\texttt{MH\_03}} & 0.050  & 0.110 & 0.096  & 0.082  & 0.125   & 0.050  & 0.035 & \textbf{0.028} \\
\textbf{\texttt{MH\_04}} & 0.100 & 0.150 & 0.197 & 0.189 & 0.280   & 0.089  & \textbf{0.051} & 0.066 \\
\textbf{\texttt{MH\_05}} & 0.080  & 0.240 & 0.206 & 0.141 & 0.284   & 0.112 & 0.082 & \textbf{0.068} \\
\midrule
\textbf{\texttt{V1\_01}} & 0.040  & 0.050  & 0.050  & 0.043  & 0.076    & 0.037  & 0.038 & \textbf{0.035} \\
\textbf{\texttt{V1\_02}} & 0.020  & 0.110 & 0.066  & 0.037  & 0.069    & 0.021  & 0.014 & \textbf{0.013} \\
\textbf{\texttt{V1\_03}} & 0.030  & 0.120 & 0.071  & 0.036  & 0.114   & 0.035  & 0.024 & \textbf{0.019} \\
\midrule
\textbf{\texttt{V2\_01}} & 0.030  & 0.070  & 0.062  & 0.044  & 0.066    & 0.036  & 0.032 & \textbf{0.023} \\
\textbf{\texttt{V2\_02}} & 0.020  & 0.100 & 0.077  & 0.044  & 0.091    & 0.024  & \textbf{0.014} & 0.015 \\
\textbf{\texttt{V2\_03}} & -    & 0.190 & 0.028  & 0.063  & 0.096    & 0.045  & 0.024 & \textbf{0.020} \\
\midrule
\textbf{Avg}  & -  & 0.119 & 0.089  & 0.071  & 0.138   & 0.048  & 0.035 & \textbf{0.031} \\
\bottomrule
\multicolumn{9}{X}{\vspace{0.0mm}\footnotesize{
$^1$ Results taken from \cite{campos2021orb}.}}\\
\multicolumn{9}{X}{\footnotesize{
$^2$ Original OKVIS VIO re-run with IMU and frontend parameters as used for OKVIS2.}}
\end{tabularx}
\caption{{EuRoC} Absoltue Trajectory Error (ATE) in [m].}
\label{t:euroc}
\end{table}

As can be seen, results on-par with or even slightly better thatn ORB-SLAM3 \cite{campos2021orb} are achieved.

\subsection{TUM VI Benchmark}
The system is also benchmarked on TUM VI \cite{schubert2018vidataset}, with respective ATE given in Table \ref{tab:tumvi}.
\begin{table}[!htb]
\setlength\tabcolsep{3.1pt} 
 \centering
 \begin{tabularx}{\columnwidth}{m{-0.2cm}c|rrrrr|rr|r|c}
  & & \rotatebox[origin=lb]{90}{\textbf{ROVIO} \cite{bloesch2017iterated}$^1$}
  & \rotatebox[origin=lb]{90}{\textbf{BASALT} \cite{usenko2019visual}$^1$} 
  & \rotatebox[origin=lb]{90}{\textbf{OKVIS} \cite{leutenegger2015keyframe} $^2$} 
  & \rotatebox[origin=lb]{90}{\textbf{OKVIS2 VIO (ours)}} 
  & \rotatebox[origin=lb]{90}{\textbf{OKVIS2 causal (ours)}}
  & \rotatebox[origin=lb]{90}{\textbf{ORB-SLAM3} \cite{campos2021orb}$^1$}
  & \rotatebox[origin=lb]{90}{\textbf{OKVIS2 (ours)}}
  & \rotatebox[origin=lb]{90}{\textbf{length [m]}} & \rotatebox[origin=lb]{90}{\textbf{loop closures}} \\
  \midrule
  & & \multicolumn{5}{c|}{Causal evaluation} & \multicolumn{2}{c|}{Non-causal} & & \\
\midrule
  & & \multicolumn{4}{c|}{VIO} & \multicolumn{3}{c|}{VI-SLAM} & & \\
\midrule
\multirow{5}{*}{\rotatebox[origin=c]{90}{\textbf{\texttt{corridor}}}} 
  & \textbf{\texttt{1}}  & 0.47 & 0.34 & 0.65 & 0.47 & {0.09} & \textbf{0.03} & {0.09} & 305 & \checkmark \\
  & \textbf{\texttt{2}}  & 0.75 & 0.42 & 0.70 & 0.49 & {0.06} & \textbf{0.02} & {0.06} & 322 & \checkmark \\
  & \textbf{\texttt{3}}  & 0.85 & 0.35 & 1.58 & 0.44 & {0.02} & \textbf{0.02} & \textbf{0.02} & 300 & \checkmark \\
  & \textbf{\texttt{4}}  & 0.13 & 0.21 & 0.15 & \textbf{0.11} & {0.18} & 0.21 & \textbf{0.18} & 114 & \\
  & \textbf{\texttt{5}}  & 2.09 & 0.37 & 0.41 & 0.47 & {0.09} & \textbf{0.01} & {0.09} & 270 & \checkmark \\
  \cmidrule(lr{0.5em}){2-11}
   &\textbf{Avg} & 0.86 &0.34 &0.70 &0.40 &{0.09} &\textbf{0.06} &{0.09} & 262 &\\
  \midrule
  \multirow{6}{*}{\rotatebox[origin=c]{90}{\textbf{\texttt{magistrale}}}} 
  &\textbf{\texttt{1}}  & 4.52  & 1.20 &2.91 &1.79 &0.09 & 0.24 &\textbf{0.03} & 918 & \checkmark \\
  &\textbf{\texttt{2}}  & 13.43 & 1.11 &2.76 &3.03 &0.52 & {0.52} &\textbf{0.03} & 561 & \checkmark \\
  &\textbf{\texttt{3}}  & 14.80 & 0.74 &1.08 &1.66 &0.93 & 1.86 &\textbf{0.93} & 566 & \\
  &\textbf{\texttt{4}}  & 39.73 & 1.58 &3.03 &3.43 &0.76 & 0.16 &\textbf{0.04} & 688 & \checkmark \\
  &\textbf{\texttt{5}}  & 3.47  & 0.60 &1.18 &1.36 &0.17 & 1.13 &\textbf{0.01} & 458 & \checkmark \\
  &\textbf{\texttt{6}}  & -     & 3.23 &2.26 &2.85 &0.63 & 0.97 &\textbf{0.63} & 771 & (\checkmark)\\
  \cmidrule(lr{0.5em}){2-11}
   &\textbf{Avg} & - &1.41 &2.20 &2.35 &0.52 &0.81 &\textbf{0.28} & 660 &\\
  \midrule
  \multirow{8}{*}{\rotatebox[origin=c]{90}{\textbf{\texttt{outdoors}}}} 
  &\textbf{\texttt{1}}  & 101.95 &255.04 & 109.75 & \textbf{22.60} & {24.92} & 32.32 & 24.92 & 2656 & \\
  &\textbf{\texttt{2}}  & 21.67  & 64.61 & 20.26 & 12.35 & \textbf{8.84} & 10.42 & \textbf{0.04} & 1601 & (\checkmark)\\
  &\textbf{\texttt{3}}  & 26.10  & 38.26 & \textbf{21.37} & 22.97 & {27.14} & 54.77 & {27.14} & 1531 & \\
  &\textbf{\texttt{4}}  & -      & 17.53 & 14.05 & {9.68} & {7.43} & 11.61 & \textbf{7.42} & 928  &\\
  &\textbf{\texttt{5}}  & 54.32  & 7.89  & 14.74 & \textbf{5.99} & {6.67} & 7.57  & {6.67} & 1168 & \checkmark \\
  &\textbf{\texttt{6}}  & 149.14 & 65.50 & 32.26 & {20.54} & 26.37 & \textbf{10.70} & 26.37 & 2045 & \\
  &\textbf{\texttt{7}}  & 49.01  & 4.07  & 13.63 & 2.92 & {0.15} & 4.58  & \textbf{0.15} & 1748 & \checkmark \\
  &\textbf{\texttt{8}}  & 36.03  & 13.53 & 28.63 & {1.20} & {0.43} & 11.02 & \textbf{0.05} & 986 &(\checkmark) \\
  \cmidrule(lr{0.5em}){2-11}
 &\textbf{Avg}  & -   &58.30  &31.83  &12.28 &{11.75} &17.87 &\textbf{11.60} & 1583 & \\
 \midrule
 \multirow{5}{*}{\rotatebox[origin=c]{90}{\textbf{\texttt{room}}}} 
  &\textbf{\texttt{1}}  & 0.16 & 0.09 & 0.08 & 0.04 & \textbf{0.01} & \textbf{0.01} & \textbf{0.01} & 146 & \checkmark \\
  &\textbf{\texttt{2}}  & 0.33 & 0.07 & 0.09 & 0.07 & \textbf{0.01} & \textbf{0.01} & \textbf{0.01} & 142 & \checkmark \\
  &\textbf{\texttt{3}}  & 0.15 & 0.13 & 0.07 & 0.05 & \textbf{0.01} & \textbf{0.01} & \textbf{0.01} & 135 & \checkmark \\
  &\textbf{\texttt{4}}  & 0.09 & 0.05 & 0.04 & 0.02 & \textbf{0.01} & \textbf{0.01} & \textbf{0.01} & 68  & \checkmark \\
  &\textbf{\texttt{5}}  & 0.12 & 0.13 & 0.08 & 0.02 & \textbf{0.01} & \textbf{0.01} & \textbf{0.01} & 131 & \checkmark \\
  &\textbf{\texttt{6}}  & 0.05 & 0.02 & 0.03 & 0.02 & \textbf{0.01} & \textbf{0.01} & \textbf{0.01} & 67  & \checkmark \\
    \cmidrule(lr{0.5em}){2-11}
   &\textbf{Avg} & 0.15 & 0.08 & 0.07 & 0.04 &\textbf{0.01} &\textbf{0.01} &\textbf{0.01} & 115 &\\
 \midrule
   \multirow{3}{*}{\rotatebox[origin=c]{90}{\textbf{\texttt{slides}}}} 
  &\textbf{\texttt{1}}  & 13.73& 0.32 & 2.55 & 1.16 & {1.47} & 0.41 & \textbf{0.10} & 289 & (\checkmark)\\
  &\textbf{\texttt{2}}  & 0.81 & 0.32 & 2.01 & 1.37 & \textbf{0.16} & 0.49 & \textbf{0.16} & 299 & (\checkmark)\\
  &\textbf{\texttt{3}}  & 4.68 & 0.89 & 2.44 & 1.62 & 1.35 & \textbf{0.47} & {1.35} & 383 & \\
  \cmidrule(lr{0.5em}){2-11}
 &\textbf{Avg} & 6.40 & 0.51 & 2.33 & 1.38 & {0.99} & \textbf{0.45} & {0.54} & 324 & \\
  \bottomrule
  \multicolumn{11}{X}{\vspace{0.0mm}\footnotesize{
$^1$ Results taken from \cite{campos2021orb}.}}\\
\multicolumn{11}{X}{\footnotesize{
$^2$ Original OKVIS VIO re-run with IMU and frontend parameters as used for OKVIS2.}}\\
\multicolumn{11}{X}{\footnotesize{
$^3$ (\checkmark) signifies loop closures only identified with OKVIS2.}}
 \end{tabularx}
 \caption{TUM-VI Absolute Trajectory Error (ATE) in [m].}
 \label{tab:tumvi}
\end{table}

As can be seen, OKVIS2 performs on-par with ORB-SLAM3 on the shorter sequences \texttt{corridor} and \texttt{room}; however, OKVIS2 clearly shows best results in the longer sequences of \texttt{magistrale} amd \texttt{slides}, as well as, perhaps most notably, in the long \texttt{outdoors} sequences. What should be noted is that especially in \texttt{magistrale} and \texttt{slides}, OKVIS2 achieves more loop closures than reported by ORB-SLAM3, denoted by (\checkmark).

Regarding \texttt{outdoors}, the approach chosen by ORB-SLAM3 to deal with slow-moving clouds is to remove all observations to landmarks further than 40 metres, whereas OKVIS2 uses CNN-based un-matching. Note that with fine-tuning to the heavily distorted fisheye images of the dataset, as well as deeper networks, there is room for improvement.

\subsection{Timings}
The per-frame timings obtained are broken down in Table~ \ref{t:timings} on EuRoC \texttt{MH\_05\_difficult}. These were measured on a PC running Ubuntu 20.04 featuring an 11$^\mathrm{th}$ Gen Intel\textsuperscript{\textregistered} {Core\texttrademark} i7-11700K with 8 cores running at up to 3.60GHz.
\begin{table}[!htb]
 \centering
 \begin{tabular}{lrp{8mm}}
\textbf{Function} & \multicolumn{2}{p{28mm}}{\textbf{Timings (std.\ dev.) [ms]}}\\
 \midrule
Detect \& Describe & $7.1$ & $(\pm1.8)$\\
Match keypoints and triangulate & $26.6$ & $(\pm12.4)$\\
Attempt loop closures & $17.7$ & $(\pm24.0)$\\
Optimise realtime graph & $33.2$ & $(\pm9.2)$\\
Pose graph edge creation \& handling & $14.0$ & $(\pm8.6)$\\
\bottomrule
\end{tabular}
\caption{Per-frame timings obtained on the EuRoC sequence \texttt{MH\_05\_difficult}. Note that the loop closure attempts are skipped in VIO mode. The synchronisation of full graph optimisations running in the background is only needed sporadically and amounts to $15.1 (\pm7.4)$ ms, when occurring. Furthermore, the detection and description is run while the realtime optimisation is still ongoing. Therefore the numbers cannot simply be summed here to check for realtime capabilities.}
\label{t:timings}
\end{table}

Note that the realtime estimator will always carry out 10 optimisation steps here. A realtime mode is also implemented that adjusts the optimisations per frame according to remaining time budget after matching and loop-closure attempts--which only minimally affects accuracy. 

Loop closure optimisations carried out in the background typically take tens of milliseconds up to around a second for the very long loops e.g.\ in the TUM-VI \texttt{outdoors} sequences.

\section{Conclusions}
A multi-camera VI-SLAM system was presented that builds a factor graph of visual reprojection errors, IMU-preintegrated error terms, as well as two-pose factors obtained from marginalisation of common observations. A realtime estimator minimises these in a bounded-size window of recent keyframes and pose-graph frames. Upon loop-closure, old pose-graph edges can seamlessly be turned back into reprojection errors and landmarks. Furthermore, longer loops can be asynchronously optimised re-using the same factor-graph whereby keeping all states around the loop as part of the optimised variables. A series of experiments demonstrates that the method achieves competitive results relative to state-of-the-art VI-SLAM systems. 

As part of future work, we may want to address robust handling of the monocular VI-SLAM case, as well as inclusion of additional sensors, such as absolute positions. Furthermore, we plan to release an open-source implementation of OKVIS2. 

Moreover, as a long-term goal, we will be exploring to what extent dense and semantic map representations can be used in a tightly-coupled manner in such a multi-sensor fusion setting, further pushing the boundaries of an integrated, robust and accurate Spatial AI solution empowering the mobile robotics and AR/VR applications of the future.






\bibliographystyle{IEEEtran}
\bibliography{./bibliography/main.bib}

\end{document}